\documentclass[fleqn,usenatbib]{mnras}

\usepackage{newtxtext,newtxmath}

\usepackage[T1]{fontenc}
\usepackage{ae,aecompl}
\usepackage{embedfile} \embedfile{mnras_template.tex}

\usepackage{graphicx}	
\usepackage{amsmath}	
\usepackage{amssymb}	
\hypersetup{final}
\usepackage{grffile}
\usepackage{multicol}

\title[high velocity collision mass loss]{Atmospheric Mass Loss from High Velocity Giant Impacts}

\author[A. Yalinewich and H. E. Schlichting]{
Almog Yalinewich,$^{1}$\thanks{E-mail: almog.yalin@gmail.com}
Hilke Schlichting,$^{2,3}$
\\
$^{1}$Canadian Institute for Theoretical Astrophysics, 60 St. George St., Toronto, ON M5S 3H8, Canada\\
$^{2}$Department of Earth, Planetary, and Space Sciences, University of California, Los Angeles, CA 90095, USA\\
$^{3}$Department of Earth, Atmospheric and Planetary Sciences, Massachusetts Institute of Technology, 77 Massachusetts Avenue, Cambridge, MA 02139, USA\\
}

\date{Accepted XXX. Received YYY; in original form ZZZ}

\pubyear{2018}

\hypersetup{draft}
\begin{document}
\label{firstpage}
\pagerange{\pageref{firstpage}--\pageref{lastpage}}
\maketitle

\begin{abstract}
Using moving mesh hydrodynamic simulations, we determine the shock propagation and resulting ground velocities for a planet hit by a high velocity impactor. We use our results to determine the atmospheric mass loss caused by the resulting ground motion due to the impact shock wave. We find that there are two distinct shock propagation regimes: In the limit in which the impactor is significantly smaller than the target ($R_i<< R_t$), the solutions are self-similar and the shock velocity at a fixed point on the target scale as $m_i^{2/3}$, where $m_i$ is the mass of the impactor. In addition, the ground velocities follow a universal profile given by $v_g/v_i=(14.2x^2-25.3x+11.3)/(x^2-2.5x+1.9) +2\ln{R_i/R_t}$, where $x=\sin\left(\theta/2\right)$, $\theta$ is the latitude on the target measured from the impact site, and $v_g$ and $v_i$ are the ground velocity and impact velocity, respectively. In contrast, in the limit in which the impactor is comparable to the size of the target ($R_i \sim R_t$), we find that shock velocities decline with the mass of the impactor significantly more weakly than $m_i^{2/3}$. We use the resulting surface velocity profiles to calculate the atmospheric mass loss for a large range of impactor masses and impact velocities and apply them to the Kepler-36 system and the Moon forming impact. Finally, we present and generalise our results in terms of the  $v_g/v_i$ and the impactor to target size ratio ($R_i/R_t$) such that they can easily be applied to other collision scenarios.
\end{abstract}




\section{Introduction}

Giant impacts are the last major assembly stage in the formation of the terrestrial planets \citep[e.g.][]{Agnor1999OnFormation,Chambers2001MakingPlanets} and may also have played a role in the formation of the close-in multiple-planet systems discovered by the {\it Kepler} satellite \citep[e.g.][]{Inamdar2014TheImpacts,Inamdar2015StealingDensities,Izidoro2017BreakingChains}.

In addition to understanding the primordial volatile budget of the Earth and any modifications due to the giant impact phase \citep[e.g.][]{Melosh1989ImpactMars,Genda2003SurvivalAspects,Schlichting2018AtmosphereLosses}, atmospheric mass loss due to giant collisions has also been invoked to explain  some of the observed diversity in exoplanet bulk densities and envelope fractions \citep[e.g.][]{Inamdar2015StealingDensities,Liu2015GIANTSUPER-EARTHS, Biersteker2018AtmosphericEnvelopes}. In particular, some of the diverse compositions in tightly packed multiple-planet systems, like, for example,  Kepler 36b and c \citep{Carter2012Kepler-36:Densities} and Kepler 11b and c \citep{Lissauer2011AKepler-11}, might be explained by atmospheric loss due to a giant collision.

 When a giant impact occurs it results in a shock wave that travels through the interior of the planet. As the shock breaks out from the surface, it moves the ground, and this motion can launch a shock into the target's atmosphere which can lead to atmospheric loss. In this work we investigate the atmospheric mass loss from giant collisions with a focus on providing a more realistic treatment of the shock's propagation through the target's interior than in previous work \citep{Schlichting2015AtmosphericImpacts}. In general, impact events and the subsequent crater formation are a complicated problem due to the myriad of physical and chemical processes involved \citep[e.g][]{Melosh2007PhysicalRim}. To simplify the problem and to gain insight into the key physical processes at play, we will assume a head on collision, a homogeneous composition and identical densities for both impactor and target. In addition, we focus here on the strong shock regime, meaning that the shock velocity is much larger than the ambient speed of sound in the target.

Many previous works used numerical simulations to study impact events \citep[e.g.][]{Melosh1992DynamicImpacts, Canup2013Lunar-formingSimulations, Chau2018FormingImpacts, Liu2015GIANTSUPER-EARTHS, Sekine2017TheRegions}. The vast majority of these were tailored to a specific impact scenario, so their results can not be easily generalised. A few works did perform parameter space surveys \cite[e.g.][]{Stewart2009VELOCITY-DEPENDENTPLANETESIMALS, Leinhardt2011CollisionsLaws, Suetsugu2018CollisionalBodies}, but these studies focused on the aftermath of the collision. In contrast, in this work we perform a parameter space survey which focuses on the shock wave's propagation through the target and we investigate the key physical difference in the behaviour of large and small impactors. We use these results to calculate the atmospheric mass loss from a wide range of impactor to target mass ratios and impact velocities.

This work extends our previous work by using numerical hydrodynamic simulations and analytic arguments to better characterise the shock's propagation through the target \citep{Schlichting2015AtmosphericImpacts}. This more realistic treatment reveals two effects that were not taken into account in \cite{Schlichting2015AtmosphericImpacts}, both of which act to increase the atmospheric mass loss. Specifically, we show that the shock wave decelerates slower than momentum conservation and that it does not decelerate considerably while the swept up mass is comparable to the mass of the impactor. The latter effect is because when the impactor is large, the rarefaction wave from the back of the impactor does not catch up to the shock wave in the target before the latter reaches the other side of the target. In this limit the mass is the hot region (i.e. in between the shock and the rarefaction wave) is comparable to the impactor mass, and hence the shock does not decelerate.

The plan of the paper is as follows. In section \ref{sec:numerical_simulations} we use hydrodynamic simulations to calculate the ground velocity as a function of position on the surface and impactor size. In section \ref{sec:atmospheric_mass_loss} we calculate the atmospheric mass loss as a function of the impactor size and velocity. In section \ref{sec:application} we use our model to estimate the atmospheric mass loss in the case of Kepler 36 and in the Moon forming impact. Finally, we discuss our results in section \ref{sec:conclusion}.

\section{Numerical Simulations} \label{sec:numerical_simulations}

\subsection{Setup}

\begin{figure}
\includegraphics[width=0.9\columnwidth]{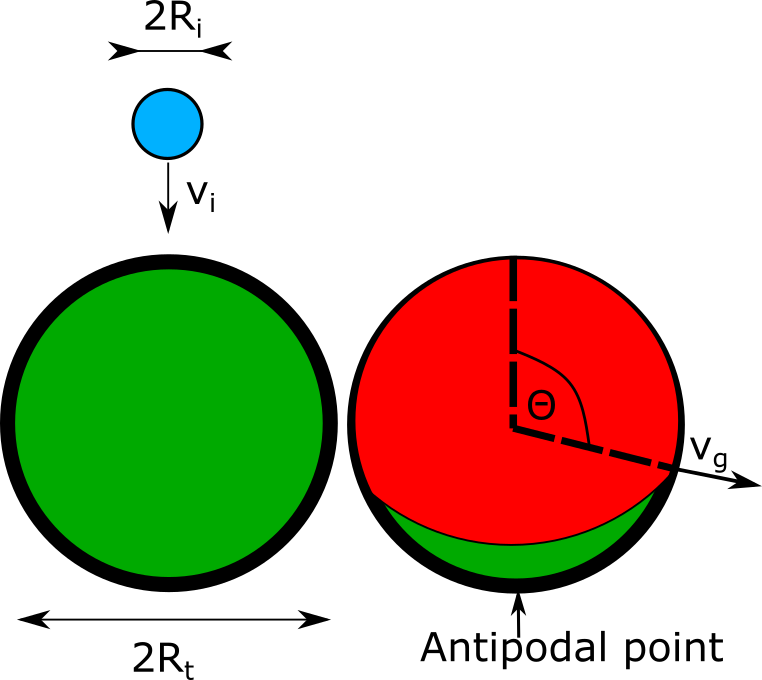}
\caption{
Schematic illustration of the simulation setup on the left, and a schematic snapshot at a later time when the shock is close to reaching the antipodal point relative to the impact site. The impactor is represented by a blue circle with radius $R_i$, and the target by a green circle with radius $R_t$. They collide with velocity $v_i$. We track the velocity, $v_a$, with which the shocked fluid (red) moves once it reaches the antipodal point. We determine (1) the ground velocity measured perpendicular to the surface, $v_g$, as a function of $\theta$, the angle between the normal to the surface of the target at a given point and the normal to the surface at the impact site, and (2) the ratio between the velocities $v_a/v_i$, where $v_a = v_g \left(\theta=\pi\right)$, as a function of the radius ratio $R_i/R_t$. 
\label{fig:giant_impact_illustration}
}
\end{figure}

We performed a series of numerical experiments in which we collided an impactor with a larger target, using the moving mesh numerical simulation RICH \citep{Yalinewich2015Rich:Mesh}. This simulation advances the hydrodynamic profiles in time by dividing the computational domain into cells, and calculate the hydrodynamic fluxes between them by solving the Riemann problem on every interface. Both target and impactor are represented as uniform density spheres, with the same density. The speed of sound in both bodies was chosen to be four orders of magnitude smaller than the collision velocity, to ensure we are in the strong shock regime. Our simulations cannot handle vacuum, so we had to fill the entire volume of the simulation with a low density gas, whose density is smaller than that of the impactor and target by nine orders of magnitude. The equation of state was assumed to follow an ideal gas
\begin{equation}
    p = \left(\gamma-1\right) \rho e
\end{equation}
where $p$ is the pressure, $\rho$ is the mass density, $e$ is the internal energy per unit mass and $\gamma$ is the adiabatic index. We assume $\gamma = 5/3$. These simulations did not include gravity. In each simulation we tracked the velocity on the surface of the target at different latitudes. A schematic illustration of the simulation setup is shown in figure \ref{fig:giant_impact_illustration}.

\subsection{Results}
\begin{figure*}
\includegraphics[width=0.7\textwidth]{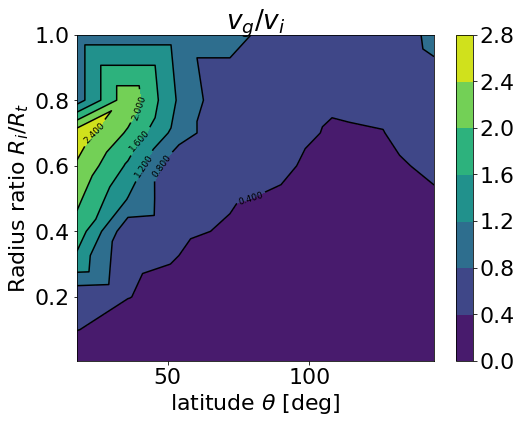}
\includegraphics[width=0.7\textwidth]{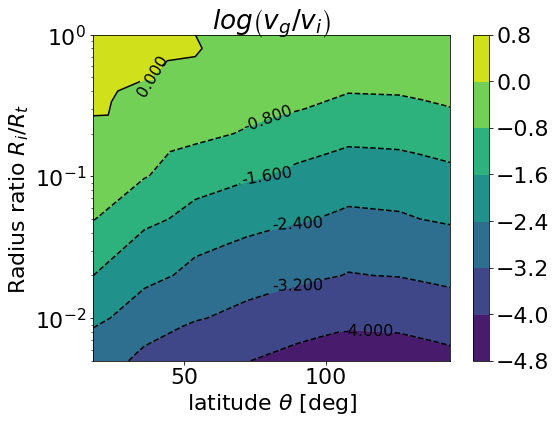}
\caption{
The ground velocity measured perpendicular to the surface of the target at different latitudes for different impactor sizes. The velocity is normalised by the impact velocity, and the radius of the impactor is normalised by the radius of the target. The latitude, radii and velocities are defined as shown in figure  \ref{fig:giant_impact_illustration}. The top panel shows the results in a linear scale, while the bottom panel shows the same results in a logarithmic scale.
\label{fig:velocity_vs_radius}
}
\end{figure*}

\begin{figure}
\includegraphics[width=0.9\columnwidth]{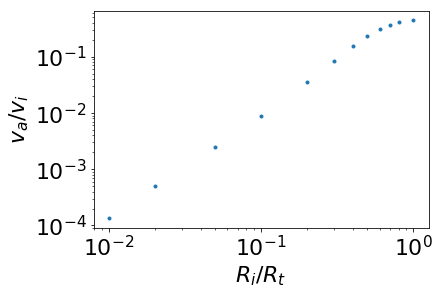}
\caption{
The ratio between the shocked fluid velocity at the antipodal point, $v_a$, and the impact velocity, $v_i$, as a function of the impactor to target ratio, $R_i/R_t$. The radii and velocities are defined and illustrated in figure \ref{fig:giant_impact_illustration}. We identify self similar behaviour of $v_a/v_i$ for small impactor to target ratios (i.e. $R_i/R_t << 1$) and an almost constant value for $v_a/v_i$ for $R_t \sim R_i$.
\label{fig:antipodal_velocity}
}
\end{figure}

A map of the ground velocity as a function of latitude and impactor size is shown in figure \ref{fig:velocity_vs_radius}. We can identify two regimes in the plot. One regime corresponds to the case when the impactor is comparable in size to the target, and the other when the impactor is much smaller. These two regimes are even more visible in figure \ref{fig:antipodal_velocity}, which shows the velocity at the antipodal point as a function of impactor size. Below we discuss the results of these two regimes in detail and show what determines the transition between them.

\begin{figure}
\includegraphics[width=0.9\columnwidth]{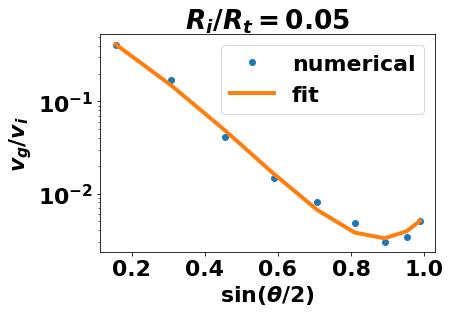}
\caption{
A comparison between the universal curve (equation \ref{eq:universal_curve}) and the numerical results for the case $R_i/R_t = 0.05$.
\label{fig:universal_curve}
}
\end{figure}

\subsubsection{Small Impactors}
In the regime where the impactor is much smaller then the radius of the target, the velocity as a function of latitude assumes, as expected, a self similar profile as displayed in figures \ref{fig:velocity_vs_radius} and \ref{fig:antipodal_velocity}. Therefore, if we know the velocity profile $v_{g} \left(\theta\right)$ (where $\theta$ is the angle between the normal at a given point on the surface and the normal at the impact site) for a certain impactor of radius $R_1$, then the velocity profile for another impactor with radius $R_2$ is simply the old velocity profile scaled by $\left(\frac{R_2}{R_1}\right)^2$. Another way to describe this power law is to say that the ground velocity at a fixed latitude scales with the impactor mass as $v_g \propto m_i^{2/3}$. This is the same scaling law as calculated in the case of crater excavation by strong explosions \citep{Zeldovich1967PhysicsPhenomena, Zahnle1995AImpact} and similar to the scaling law observed in nuclear explosions \citep{Perret1975Free-fieldExplosions}. Using this scaling law, we can calibrate a universal curve for the ground velocity perpendicular to the surface of the target, $v_g$, given by
\begin{equation}
    \ln \frac{v_g}{v_i} = \frac{14.2 x^2-25.3 x+11.3}{1.0 x^2-2.5 x+1.9} + 2 \ln \frac{R_i}{R_t} \label{eq:universal_curve}
\end{equation}
where $x = \sin \frac{\theta}{2}$. A comparison between this fit and the numerical results is presented in figure \ref{fig:universal_curve}. 

\subsubsection{Transition between Small and Large Impacts}
\begin{figure}
\includegraphics[width=0.9\columnwidth]{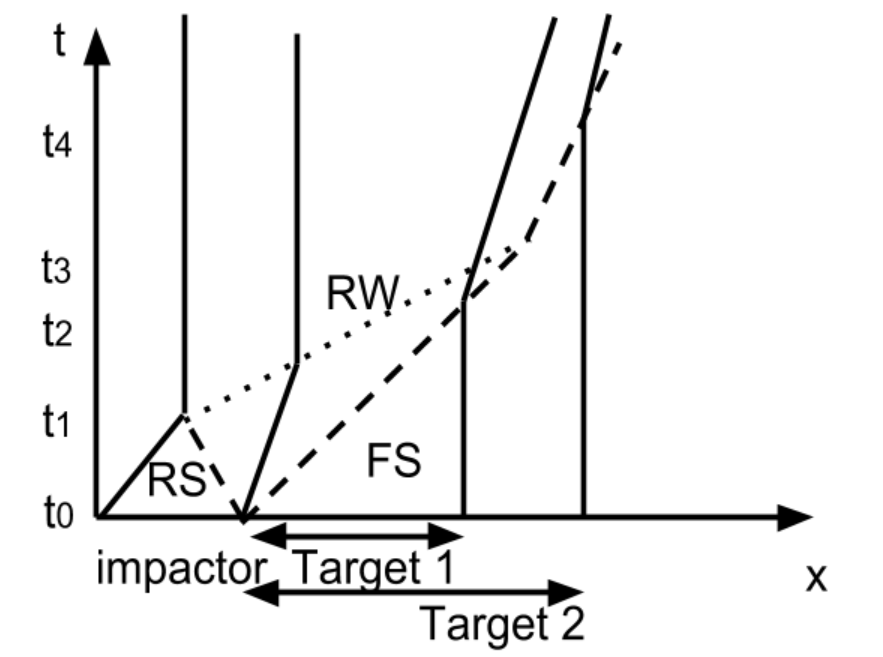}
\caption{
Schematic wave trajectories during a collision. Solid lines represent boundaries of the two bodies, dashed lines represent shock waves and the dotted line is the rarefaction wave. We show two target thicknesses (denoted by target 1 and 2) to demonstrate what happens if shock breakout occurs before and after the overtaking by the rarefaction wave. The meaning of the acronyms is as follows: FS - forward shock, RS - reverse shock and RW - rarefaction wave.
\label{fig:hydro_traj}
}
\end{figure}

\begin{figure}
\includegraphics[width=0.9\columnwidth]{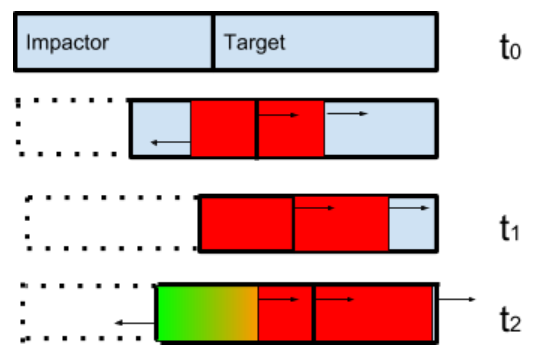}
\caption{
Schematic time sequence (from top to bottom) showing the motion of waves inside the target and impactor in a setting where the forward shock reaches the antipodal point of the target before being overtaken by the rarefaction wave. Blue represents the unshocked material, red is the hot region and the green - orange gradient represents the rarefaction wave.
\label{fig:breakout_before_overtaking}
}
\end{figure}

\begin{figure}
\includegraphics[width=0.9\columnwidth]{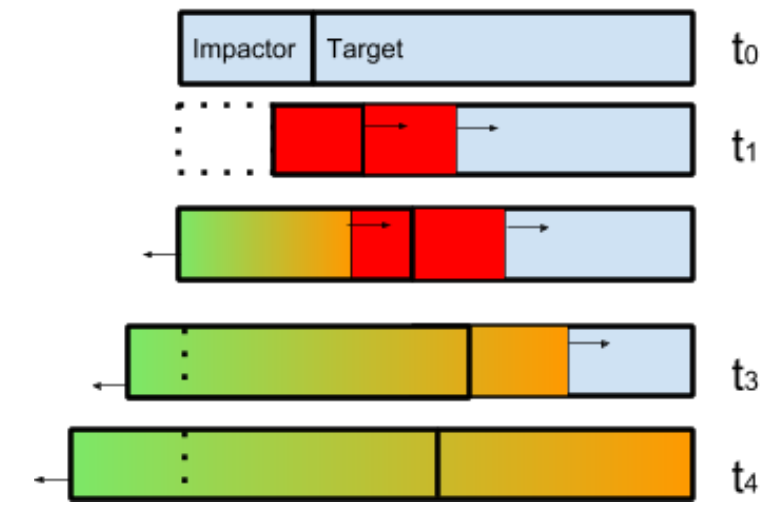}
\caption{
Schematic time sequence (from top to bottom) showing the motion of waves inside the target and impactor in a setting where the rarefaction wave overtakes the shock before breakout. Blue represents the unshocked material, red is the hot region and the green - orange gradient represents the rarefaction wave.
\label{fig:overtaking_before_breakout}
}
\end{figure}

When the impactor is comparable in size to the target, the velocity profile is different from the self similar profile shown in figure 4 and no longer described  by Equation \ref{eq:universal_curve}. To understand where the transition between the two regimes comes from, let us consider in more detail what happens in a one dimensional, slab symmetric collision. In such a collision (see figures \ref{fig:hydro_traj}, \ref{fig:breakout_before_overtaking} and \ref{fig:overtaking_before_breakout}), a forward shock forms in the target, but also a reverse shock in the impactor (see time $t_0$ in figures \ref{fig:hydro_traj}, \ref{fig:breakout_before_overtaking} and \ref{fig:overtaking_before_breakout}). Because the impactor is smaller than the target, the reverse shock traverses the impactor before the forward shock traverses the target. When the reverse shock reaches the other side of the impactor, it disappears and a rarefaction wave forms and travels towards the target (corresponding to time $t_1$ in figures \ref{fig:hydro_traj}, \ref{fig:breakout_before_overtaking} and \ref{fig:overtaking_before_breakout}). The rarefaction wave moves faster than the shock. This is because the rarefaction waves move at the speed of sound in the hot region (i.e. material that has been shocked but not cooled by the rarefaction wave) relative to the material and the shock waves move subsonically with respect to the downstream speed of sound \citep{Landau1987FluidMechanics}. For large enough impactors, the forward shock reaches the other side of the target before the rarefaction catches up to it (corresponding to time $t_2$ in figures \ref{fig:hydro_traj} and \ref{fig:breakout_before_overtaking}). For small impactors, the rarefaction wave catches up to the shock wave and weakens it (corresponding to time $t_3$ in figures \ref{fig:hydro_traj} and \ref{fig:overtaking_before_breakout}) before the forward shock reaches the other side of the target. The rarefaction wave can be thought of as being composed of multiple wavelets of declining pressures and sound speeds. Each time another wavelet makes it to the shock front, the pressure decreases as the shock decelerates further. The hydrodynamic trajectories are illustrated in figure \ref{fig:hydro_traj}. The motion of the waves in the case where the forward shock reaches the antipodal point before it is overtaken by the rarefaction wave is illustrated in figure \ref{fig:breakout_before_overtaking}, and the case where it is overtaken in figure \ref{fig:overtaking_before_breakout}. The same trajectories are also illustrated in a series of one dimensional numerical simulations presented in appendix \ref{app:1d}.

In a purely one dimensional, slab symmetric collisions between shells the critical ratio between the size of the impactor and target can be obtained analytically. The velocity of the hot material immediately after the impact is $v_s = \frac{1}{2} v_i$. The velocity of the shock is $V_s = \frac{\gamma+1}{4} v_i$. The time it takes the shock to cross the target is $t_2 = 2 R_t / V_s = \frac{4}{\gamma+1} \frac{2 R_t}{v_i}$. The time it takes the shock to cross the impactor is $t_1 = 2 R_i/V_s = \frac{4}{\gamma+1} \frac{2 R_i}{v_i}$. When the shock crosses the impactor, it is reflected from the free surface and moves with the shocked speed of sound $c_s = \sqrt{\frac{\gamma \left(\gamma-1\right)}{8}} v_i$ relative to the hot material. The rarefaction wave travels through the impactor, and then the target, which have been compressed by a factor of $\left(\gamma+1\right)/\left(\gamma-1\right)$. The time it takes the rarefaction wave to traverse the impactor and target, if it does not overtake the shock, is $t_3 = \frac{\sqrt{8 \left(\gamma-1\right)}}{\left(\gamma+1\right) \sqrt{\gamma}} \frac{2 R_i + 2 R_t}{v_i}$. Hence, the rarefaction wave overtakes the shock when $t_2 > t_1+t_3$, which yields
\begin{equation}
    \frac{R_i}{R_t} < \frac{2 \sqrt{\gamma}-\sqrt{2 \left(\gamma-1\right)}}{2 \sqrt{\gamma}+\sqrt{2 \left(\gamma-1\right)}}.
\end{equation}
Equation 2 evaluates to $\sim 0.4$ for $\gamma =5/3$, which is in agreement with equivalent one dimensional slab simulations that we performed (not shown in this work).

\subsubsection{Large Impactors} \label{sec:large_impactors}

We see in figure \ref{fig:antipodal_velocity}
that the velocity ratio, $v_a/v_i$, becomes almost constant for impactors that are comparable in size to the target (i.e. $R_i/R_t \sim 1$). This is because the information about the finite size of the impactor is carried by the rarefaction wave that is reflected from the back side of the impactor, and for wide enough impactors the forward shock traverses the target before the rarefaction wave catches up to it. As long as the rarefaction wave can't catch up with the forward shock (i.e. in the large impactor regime), the velocity of the shock remains close to constant. See appendix A1 for details.

Another way to understand why the shock velocity is almost constant is to consider the momentum budget. Right after the impact, a forward shock emerges from the point of contact and travels inside the target. This shock wave sweeps up more mass from the target, but does not decelerate because the impactor keeps transferring momentum. After the shock makes it to the back side of the impactor, it is reflected as a rarefaction wave. After that, for every amount of mass swept up by the shock front, a larger amount of mass is swept by the rarefaction wave. The shock wave compresses, heats up and accelerates the swept up material, while the rarefaction wave does the opposite. In terms of the momentum budget, one can think of it as though material is accelerated by the shock at the expense of material decelerated by the rarefaction wave. The rarefaction wave first traverses the impactor, seeping out the remaining momentum, and then continues to propagate in the target. The rarefaction wave then ``eats'' into the hot region behind the shock. We use here the term ``hot" to distinguish this region from the cold upstream and material that has already cooled by the rarefaction wave. The wave trajectories described above are illustrated in detail in appendix \ref{app:1d}.

This behaviour is also in someways similar to Newton's cradle, a toy made up of several juxtaposed metal balls suspended by wires. When a ball on one end is pulled and released, it swings and hits the next ball, but only the ball at the other end moves as a result of the impact with almost the same amplitude as the first. This is because the collision gives rise to a rectangular sound wave that travels through the row of balls without attenuation and without dispersion since, in the sub sonic regime, both the shock and the rarefaction wave travel at a velocity close to the speed of sound. As a result, the rarefaction wave cannot catch up to the shock. To illustrate this behaviour, we simulated a slab symmetric, one dimensional collision at a sub sonic velocity, and the results are shown in appendix \ref{app:1d}3.

In spherical collisions, there is a decrease of the velocity ratio, $v_a/v_i$, with decreasing impactor to target ratios, $R_i/R_t$, because of the geometry of the problem. As mentioned above, when a plane shock wave reaches the edge of a slab, it is reflected as a rarefaction wave. As a result of this reflection, the pressure in the shocked region drops to zero, and the shocked fluid accelerates further. In a spherical collision, due to the geometry, lateral rarefaction waves trail after the shock wave and suppress the shock reflection.


\section{Atmospheric Mass Loss} \label{sec:atmospheric_mass_loss}
\begin{figure*}
    \includegraphics[width=0.7\linewidth]{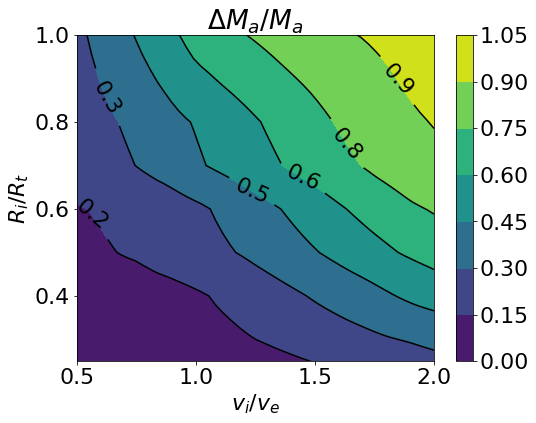}
    \includegraphics[width=0.7\linewidth]{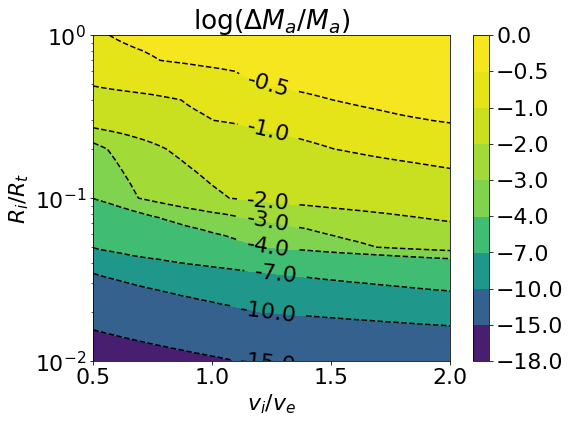}
\caption{Fraction of the atmosphere lost as a result of a giant collision, as a function of the mass and velocity of the impactor. The radius of the impactor is normalised by the radius of the target, and the velocity is normalised by the escape velocity. The bottom panel zooms in on the small impactor to target size regime and displays the mass loss in a logarithmic scale.}
\label{fig:atmosphere_loss_linear}
\end{figure*}

Using the results from section \ref{sec:numerical_simulations}, we calculate the atmospheric mass loss for each gas column above a given point on the surface of the target. If the ground velocity is larger than the escape velocity then the whole atmospheric column above that point is lost. If, however, the ground velocity is lower than the escape velocity on the surface only the upper parts of the atmosphere are lost. This is due to the fact that the density in the atmosphere decreases with altitude, which causes the shock wave to accelerate with height in the atmosphere. The fraction of lost atmosphere as a function of the ratio between ground and escape velocity $\chi_{\rm loss} \left(v_g/v_e\right)$ was calculated in  \citet{Schlichting2015AtmosphericImpacts} for different kinds of atmospheres. For simplicity, we assume an adiabatic atmosphere with adiabatic index $\gamma=5/3$ \citep[see figure 5 in][]{Schlichting2015AtmosphericImpacts}. The total relative atmospheric mass loss is given by
\begin{equation}
    \frac{\Delta M_a}{M_a} = \frac{1}{2 R_t^2}\int_0^{2 R_t} \chi_{\rm loss} \left(v_g/v_e\right) l dl,
\end{equation}
where $l = 2 R_t \left(1 - \cos \left(\theta/2\right)\right)$. Using the atmospheric mass loss results from \citet{Schlichting2015AtmosphericImpacts}  together with the refined distribution of ground velocities over the planet's surface from this work, we calculate the atmospheric mass loss for various impactor sizes and a range of impact velocities. Figure \ref{fig:atmosphere_loss_linear} displays a two dimensional map of the relative atmospheric mass loss as a function of the mass ratio between the impactor and target, and the ratio between the impact velocity and escape velocity. 

The mass loss predicted in this work exceeds that in \citet{Schlichting2015AtmosphericImpacts} for the same impact conditions due to our improved treatment of the shock propagation in the core of the target. When the impactor is comparable in size to the target the major difference is due to the ``Newton's cradle'' effect discussed in section \ref{sec:numerical_simulations}. For impactors much smaller than the target the major difference is due to the fact that the velocity at a fixed point on the target scales with the impactor mass as $v_g \propto m_i^{2/3}$, whereas \citet{Schlichting2015AtmosphericImpacts} assumed that it scales as $v_g \propto m_i$, which leads in a different distribution and smaller magnitudes of the ground velocities over the target's surface and hence less atmospheric loss.

We note that the calculation here does not take into account additional mass loss above the tangent plane of the impact site \citep{Schlichting2015AtmosphericImpacts}, which is expected to dominate the mass loss for small impactors.

\section{Applications} \label{sec:application}
Giant impacts are the last major assembly stage in the formation of the terrestrial planets \citep{Chambers2001MakingPlanets} and they are likely also important in the formation of the close-in multiple-planet systems discovered by the {\it Kepler} satellite  \citep{Inamdar2014TheImpacts,Inamdar2015StealingDensities,Izidoro2017BreakingChains}.

In this section we consider two specific collision scenarios and calculate the resulting mass loss.

\subsection{Kepler-36}

Several planets residing in multiple-planet systems show surprising diversity in their bulk densities which is hard to explain from gas accretion and subsequent thermal evolution and loss alone \citep[e.g][]{Schlichting2018FormationSuper-Earths}. One such system is Kepler-36 which hosts two planets comparable in mass on adjacent orbits (P=13.8 and 16.2 days) \citep{Carter2012Kepler-36:Densities}. The planets are so close, in fact, that they are in a 29:34 resonance \citep{Deck2012RAPIDSYSTEM}. Kepler 36b has a mass of about 4 $M_{\oplus}$ and a radius of about 1.5 $R_{\oplus}$, which is consistent with an Earth-like composition. In contrast, its neighbour Kepler 36c has a mass 8 $M_{\oplus}$ and radius 3.7 $R_{\oplus}$, which is consistent with an extended H/He envelope \citep{Lopez2013TheDichotomy}. Planet formation scenarios predict that Kepler 36b should also have formed with an extended H/He atmosphere \citep[e.g.][]{Ginzburg2015Super-EarthRetention}. One possible explanation for the lack of an extended H/He atmosphere for Kepler 36b is that it was lost in a giant impact \citep{Liu2015GIANTSUPER-EARTHS,Inamdar2015StealingDensities}.

Following \citet{Liu2015GIANTSUPER-EARTHS}, we consider a collision between a $1 M_{\oplus}$ impactor and a $4 M_{\oplus}$ target. From figure \ref{fig:atmosphere_loss_linear} we see that our model predicts an atmospheric mass loss from the shock propagation of about 40 \%. This result is slightly larger than both the one dimensional shock calculation in \citet{Inamdar2015StealingDensities} and the three dimensional shock calculation of \citet{Liu2015GIANTSUPER-EARTHS} (both estimating the atmospheric mass at 30 \%). The actual atmospheric loss from such an impact may, however, be significantly more than what is calculated here from the shock component alone. \citet{Biersteker2018AtmosphericEnvelopes} find that thermal impact heating should be able to unbind the entire hydrogen and helium atmosphere of Kepler 36b for the impact parameters investigated here. 

We note that one major difference between our model and that of \citet{Liu2015GIANTSUPER-EARTHS} is the interior model used for the impactor and target. Whereas we assume a homogeneous, constant density interior for both impactor and target, \citet{Liu2015GIANTSUPER-EARTHS} assumes that both are differentiated. Although exoplanets, like the ones found in the Kepler-36 system, are expected to differentiate eventually, it is unlikely that they will have done so by the giant impact stage. Most giant impacts in close-in exoplanet systems occur on 1 million to 100 million year timescales \citep{Izidoro2017BreakingChains}. In contrast, it takes of the order of a Gyr for the cores of super-earths to cool sufficiently for core-formation and differentiation to take place \citep[e.g.][]{Ginzburg2017Core-poweredExoplanets}. This is because the extended H/He envelope acts as a blanket providing a bottleneck for the cooling of the envelope and the underlying core \citep[e.g.][]{Lee2015ToAtmospheres,Ginzburg2015Super-EarthRetention}. We therefore believe that the simple undifferentiated model for the target and impactor that is investigated here is appropriate for the Kepler-36 system. 

\subsection{Moon Forming Collision}

It is generally believed that the Earth's Moon formed as a result of a single \citep[e.g.][]{Hartmann1975Satellite-sizedOrigin} or multiple \citep{Rufu2017AMoon} giant impacts. The canonical Moon forming impact consists of a Mars-sized impactor that collides with proto-Earth at their mutual escape velocity  \citep{Canup2001OriginFormation}. Previous works estimating the atmospheric mass loss from such an impact find typical results ranging from 5\% to 30\% \citep{Genda2003SurvivalAspects,Schlichting2015AtmosphericImpacts,Lock2014WasImpact}. We find according to our results in figure \ref{fig:atmosphere_loss_linear}, that a collision between proto-Earth and a Mars-sized object would expel 20\% of the atmosphere, in accordance with previous results.

\section{Discussion \& Conclusions} \label{sec:conclusion}

Using a moving mesh hydrodynamic simulation, we determine the shock propagation and resulting ground velocities for a planet hit by a high velocity impactor. This work builds upon a previous work \citep{Schlichting2015AtmosphericImpacts} and provides a more accurate description of the passage of the shock through the target's interior.
We find that there are two distinct shock propagation regimes: In the limit in which the impactor is significantly smaller than the target ($R_i<< R_t$), the solutions are self-similar, the shock velocity at a fixed point on the target decreases according to  $v_g \propto m_i^{2/3}$, where $m_i$ is the mass of the impactor, and the ground velocities follow a universal profile given by $v_g/v_i=(14.2x^2-25.3x+11.3)/(x^2-2.5x+1.9) +2\ln{R_i/R_t}$, where $x=\sin{\theta/2}$ and $\theta$ is the latitude measured relative to the impact site. The simple scaling laws that we obtain are a result of the self-similar nature of the shock propagation in the strong shock regime. This self-similarity  breaks down when the impactor becomes comparable in size to the target ($R_i \sim R_t$) in which case we find that shock velocities decline significantly more weakly than $m_i^{2/3}$. This weaker decline is due to the fact that the rarefaction wave trailing after the shock in the target does not catch up to it before the shock reaches the antipodal point. This behaviour is similar to Newton's cradle (see section \ref{sec:large_impactors}).

We present our ground velocity results as a function of latitude on the target and the impactor to target ratio such that they can easily be applied to different impact scenarios by members of the scientific community. In addition, we use our  ground velocity results to calculate the resulting atmospheric mass loss. To translate the global ground velocities to the total atmospheric loss by an impact shock, we use the results from \citet{Schlichting2015AtmosphericImpacts} and present our atmospheric loss results as a function of impact velocity and the size ratio of the impactor and target. We find in general more atmospheric loss due to large collisions compared to \citet{Schlichting2015AtmosphericImpacts}. This difference is entirely due to the improved treatment of the shock propagation in the target, which yields more accurate ground velocities across the surface of the target and which in turn provides more accurate initial conditions for the shock that is launched into the atmosphere and hence the atmospheric loss calculations. The main reason for the difference in the global ground velocities found in this work and used in \citet{Schlichting2015AtmosphericImpacts} is that \citet{Schlichting2015AtmosphericImpacts} assumed momentum conservation, which implies a rapid decline of the shock velocity $v_g\propto m_i$ (where $m_i$ is the mass swept up by the shock), whereas in this study we found that the velocity actually declines slower, namely $v_g \propto m_i^{2/3}$, or even weaker than this when the impactor is comparable in size to the target.

We applied our model to two impact scenarios. The first involves Kepler 36b, in which case we want to know if a giant impact could have stripped it of its gaseous envelope. We find that the shock launched from an impactor with a mass of $1 M_{\oplus}$ moving at the escape velocity can remove about a third of the atmosphere. This is in agreement with the numerical simulation in \citet{Liu2015GIANTSUPER-EARTHS}. The actual atmospheric loss from such an impact may, however, be significantly more than what is calculated here from the shock alone. \citet{Biersteker2018AtmosphericEnvelopes} find that thermal impact heating should be able to unbind the entire atmosphere of Kepler 36b for the impact parameters investigated here.

The second impact scenario to which we applied our results is the Moon forming impact, in which case we are interested in the atmospheric loss from the proto - Earth. In the case of the Moon forming impact, we find that a head on collision with a Mars-sized object at the mutual escape velocity can remove about 20\% of the atmosphere.

In this work we assumed the target is made up of an ideal gas, whereas in reality it would have a more complicated equation of state. We argue that this simplification is justified in the limit of a strong shock. In shock experiments, many materials exhibit a linear relation between the shock velocity $U_s$ and the material velocity $U_m$, just like an ideal gas. The equivalent adiabatic index can be deduced from the slope of the relation between the shock and material velocity, using the Rankine-Hugoniot conditions
\begin{equation}
    \frac{d U_s}{d U_m} = \frac{\gamma+1}{2}.
\end{equation}
For example, in the case of silica $d U_s/d U_m = 1.2$ \citep{McCoy2016Shock-wave1600GPa}, which corresponds to an equivalent adiabatic index of $\gamma = 1.4$.

Most objects hit planets close to their mutual escape velocity \citep[e.g.][]{Schlichting2014FormationImplications}, which is often comparable to the speed of sound in the planet's interior, so the assumption of a strong shock may not be fully applicable in all giant impacts. There are, however, cases when this assumption is well justified. One scenario is a small planet close to its host star, such that the keplerian velocity is larger than the planet's escape velocity. Another scenario is an impact on a planet's satellite. In both cases, the target is deep inside a potential well of another object. In addition, collisions between planets with large eccentricities, which can, for example, be produced by the Kozai-Lidov mechanism, could result in high velocity impacts that would be in the strong shock regime \citep[e.g.][]{Denham2018HiddenCompanion}.

Finally, in this work the calculations that follow the shock propagation through the target do not include gravity or the finite speed of sound in the planet. We expect the influence of gravity to be small, as most of the target mass is hardly displaced, so little kinetic energy is converted into potential energy. However, when gravity is taken into account, then it has to be balanced by a thermal pressure, which introduces a finite speed of sound to the target and which has to be of the same order of magnitude as the escape velocity. Including a finite speed of sound in the target allows for the possibility that a shock wave initially moves super sonically, but as it sweeps up more material it slows down to the point where its velocity is comparable to the speed of sound. We intend to explore this trans-sonic regime and the weak shock regime in future work.

\section*{Acknowledgements}
We thank the reviewer for useful comments that help to clarify the manuscript.
AY would like to thank Re'em Sari, Wei Zhu, Dan Tamayo and Lauren Weiss for useful discussions.
HES gratefully acknowledges support from the National Aeronautics and Space Administration under grant No. $17~\rm{NAI}18\_~2-0029$ issued through the NExSS Program. In this work we use the NumPy \citep{Oliphant2006ANumPy} and matplotlib \citep{Hunter2007Matplotlib:Environment} python packages.




\bibliographystyle{mnras}
\bibliography{references} 




\appendix

\section{One Dimensional Simulations} \label{app:1d}

In this section we present one dimensional simulations aimed at elucidating the points made in section \ref{sec:large_impactors}. More specifically, we want to illustrate the shock trajectories for different regimes discussed in that section. These simulations were carried out using a one dimensional version of RICH \citep{Yalinewich2015Rich:Mesh}, on a computational domain bounded between -2 and 2, with 1000 computational cells.

\subsection{Large Impactors}

\begin{figure}
\includegraphics[width=0.9\columnwidth]{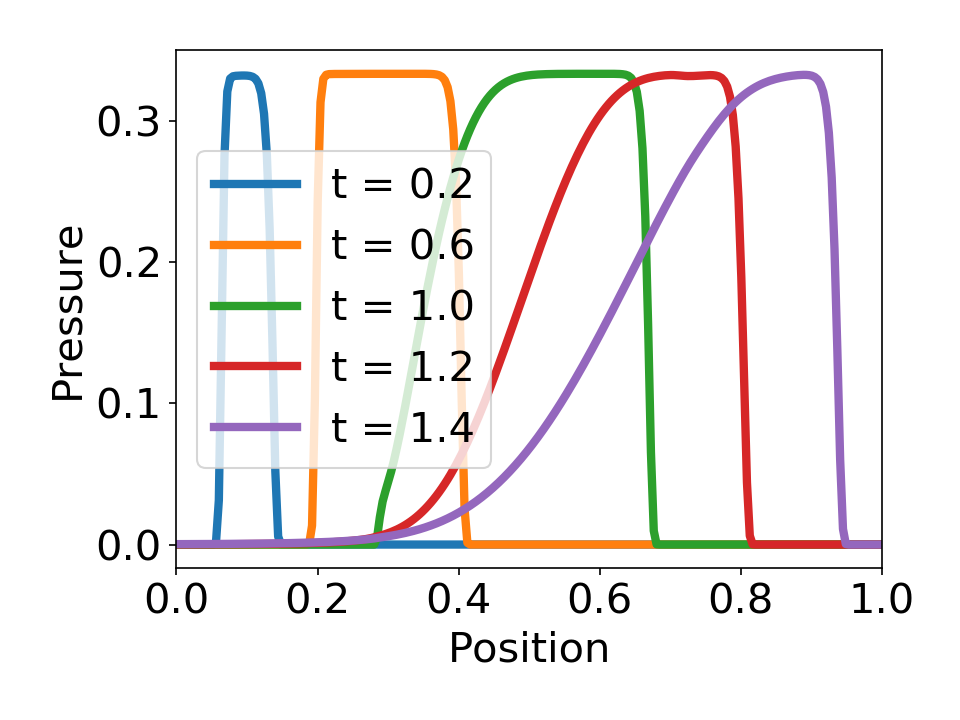}
\includegraphics[width=0.9\columnwidth]{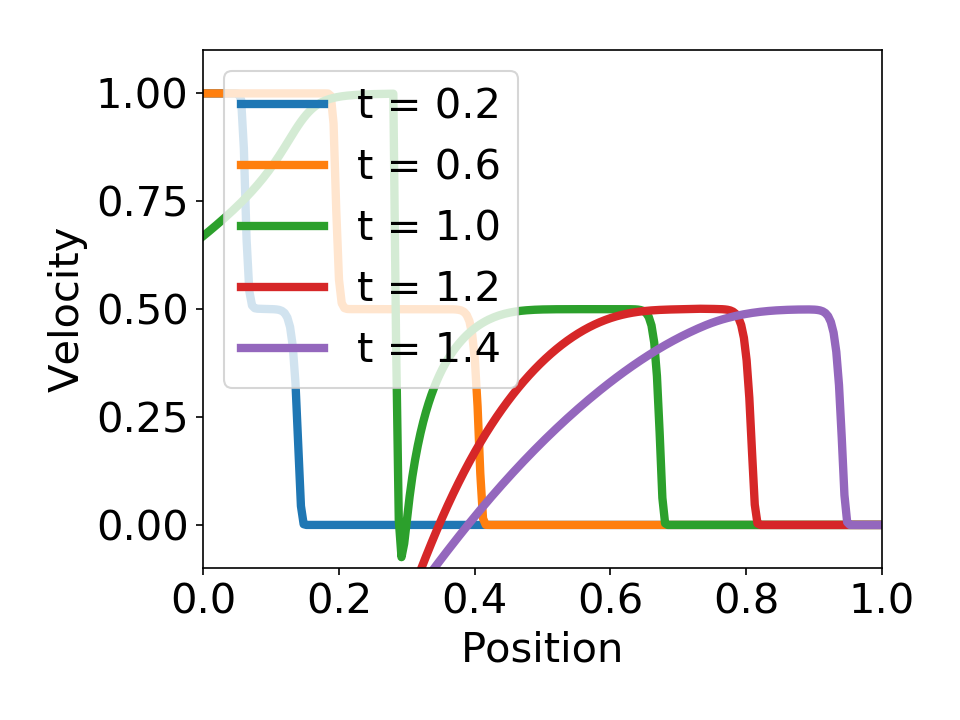}
\caption{
Numerical pressure and velocity profiles at different times due to a collision between cold slabs. The target, initially between positions 0 and 1, is hit on the left by an impactor whose width is 0.6 of the target, moving at velocity 1. Time is measured in units where 1 is the time it takes the unimpeded impactor to travels a distance equal to the thickness of the target. After time 1, one can see a rarefaction wave trailing after the shock wave and catching up to it.
\label{fig:thick_impactor_lapse}
}
\end{figure}

We simulated a collision with two bodies whose thickness ratio is 0.6. The velocity of the collision is about three orders of magnitude larger than the ambient speed of sound. Plotted in figure \ref{fig:thick_impactor_lapse} are the pressure and velocity profiles inside the target at different times. At first, the size of the hot region (i.e. material that has been swept by the shock but not by the rarefaction wave, represented in the figure by the pressure plateau) increases, but after the rarefaction wave emerges, it starts catching up to the shock wave, and the hot region shrinks with time. During this time, however, the shock wave does not decelerate. This phase is similar to Newton's cradle discussed in section A3.

\subsection{Small Impactors}

\begin{figure}
\includegraphics[width=0.9\columnwidth]{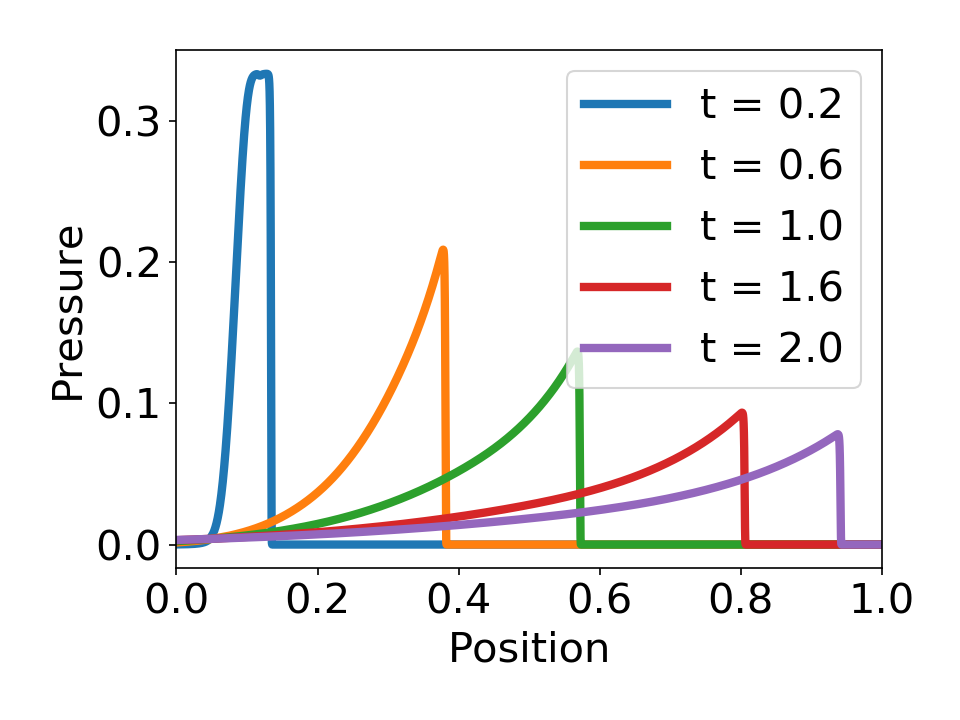}
\includegraphics[width=0.9\columnwidth]{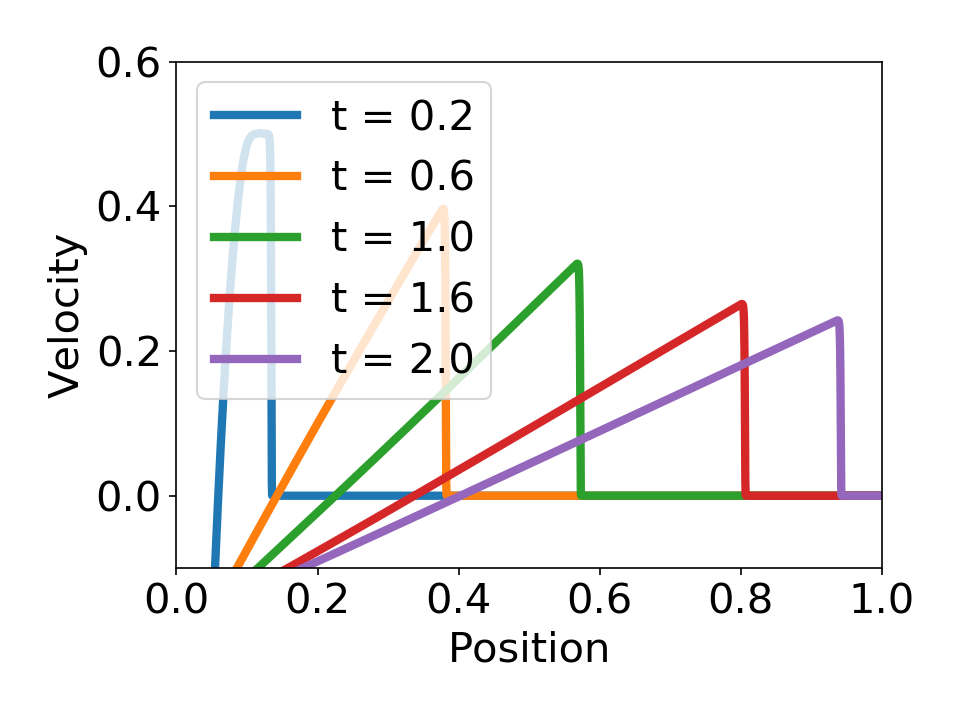}
\caption{
Numerical pressure and velocity profiles at different times due to a collision between cold slabs. The target, initially between positions 0 and 1, is hit on the left by an impactor whose width is 0.2 of the target, moving at velocity 1. Time is measured in units where 1 is the time it takes the unimpeded impactor to travels a distance equal to the thickness of the target. In this case the shock wave weakens and decelerates considerably.
\label{fig:thin_impactor_lapse}
}
\end{figure}

We simulated a collision with two bodies whose thickness ratio is 0.2. The velocity of the collision is about three orders of magnitude larger than the ambient speed of sound. Plotted in figure \ref{fig:thin_impactor_lapse} are the pressure and velocity profiles inside the target at different times. In this case the rarefaction wave quickly catches up to the shock, and the latter begins to decelerate as it sweeps more material.

To understand why the shock wave decelerates when it is caught up by the rarefaction wave, one can think about the rarefaction wave as a sequence of discrete ``wavelets", each one with a slightly lower pressure, and travelling slightly slower then the previous one. Every time each one of these wavelets reaches the shock front, it decreases the pressure there. The longer the target with respect to the impactor, the more of these wavelets reach the shock front and the slower it gets.

\subsection{Newton's Cradle}

In this section we present the results of a simulation of Newton's cradle, which was discussed in section \ref{sec:large_impactors}. We simulated a collision with two bodies whose thickness ratio is 0.2. The velocity of the collision is a factor of nine smaller than the ambient speed of sound. Plotted below in figure \ref{fig:newtons_cradle} are the pressure and velocity profiles inside the target at different times. In this case both rarefaction and shock waves travel at a velocity close to the speed of sound, so the rarefaction does not catch up to the shock.

\begin{figure}
\includegraphics[width=0.9\columnwidth]{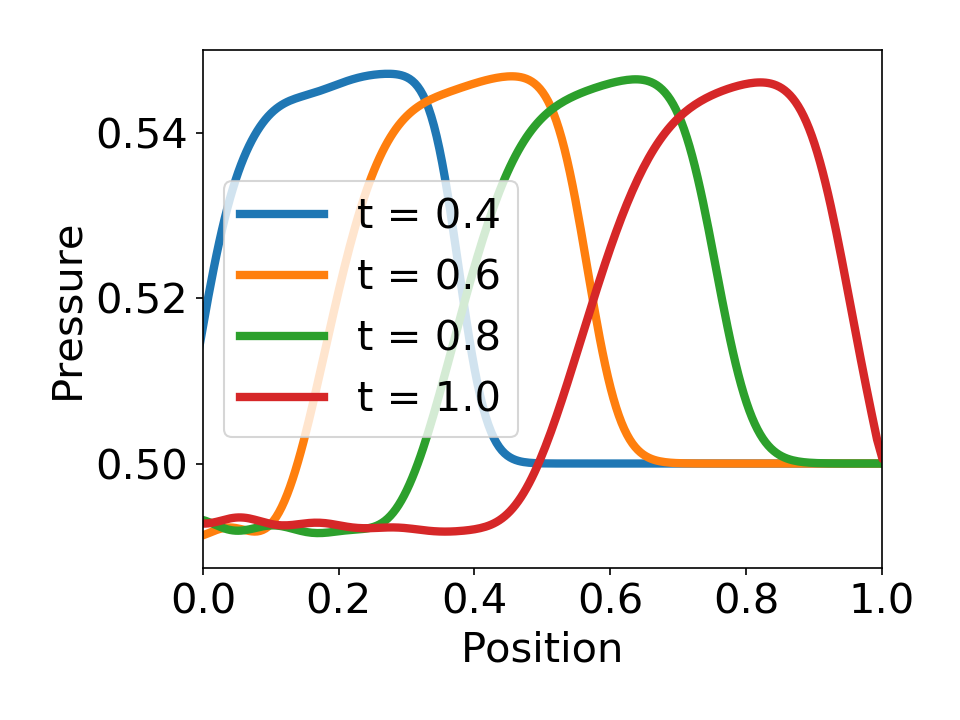}
\includegraphics[width=0.9\columnwidth]{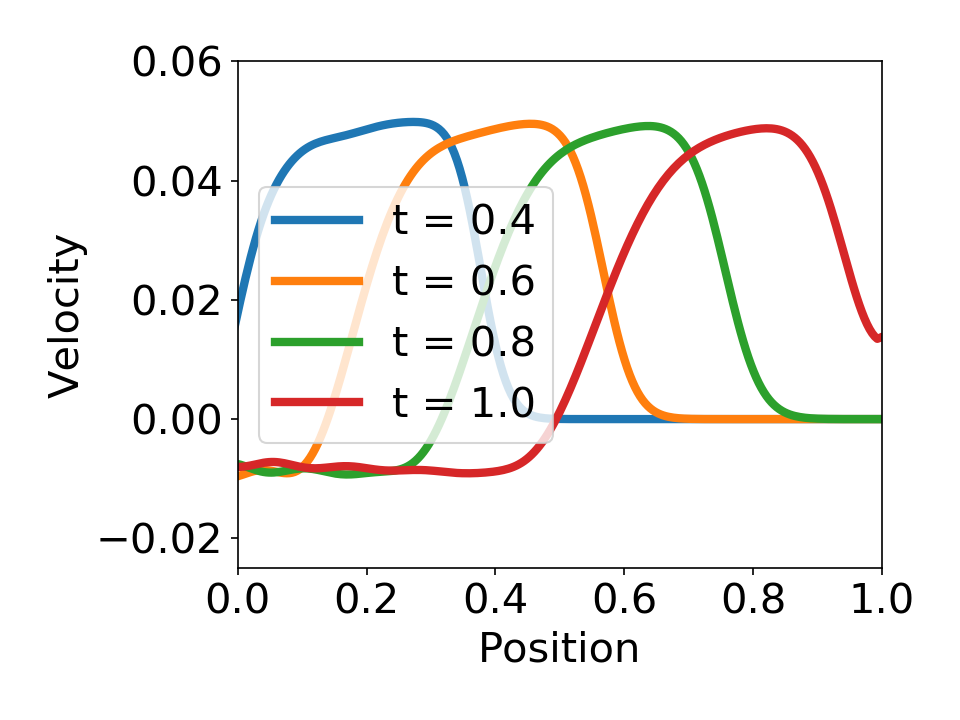}
\caption{
Numerical pressure and velocity profiles at different times due to a collision between warm slabs. The target, initially between positions 0 and 1, is hit on the left by an impactor whose width is 0.2 of the target, moving at velocity 0.1. The speed of sound in the target and impactor is 0.9. Time is measured in units where 10 is the time it takes the unimpeded impactor to travels a distance equal to the thickness of the target. In this case both rarefaction and shock waves are moving at velocities close to the speed of sound, so the pulse preserves its shape as it propagates through the target.
\label{fig:newtons_cradle}
}
\end{figure}



\bsp	
\label{lastpage}
\end{document}